\documentclass[prl,twocolumn,longbibliography,superscriptaddress]{revtex4-2}
\usepackage{graphicx}
\usepackage{bm}
\usepackage{bbm}
\usepackage{wrapfig}
\usepackage{color}
\usepackage{natbib}
\usepackage{amssymb}
\usepackage{amsmath}
\usepackage{geometry}
\usepackage{comment}
\usepackage{hyperref}

\usepackage[capitalize]{cleveref} 
\usepackage{braket} 
\usepackage{mathtools} 
\usepackage{siunitx}
\usepackage{appendix}
\usepackage{cancel}

\renewcommand{\Re}{\operatorname{Re}}

\DeclareMathOperator{\tr}{tr}
\DeclarePairedDelimiter\abs{\lvert}{\rvert}

\begin{document}
\title{Spin-valley Silin modes in graphene with substrate-induced spin-orbit coupling}

\author{Zachary M. Raines}
\affiliation{Department of Physics, Yale University, New Haven, CT 06520, USA}
\author{Dmitrii L. Maslov}
\affiliation{Department of Physics, University of Florida, Gainesville, FL, 32611, USA}
\author{Leonid I. Glazman}
\affiliation{Department of Physics, Yale University, New Haven, CT 06520, USA}

\date{\today}

\begin{abstract}
In the presence of external magnetic field the Fermi-liquid state supports oscillatory spin modes known as Silin modes.
We predict the existence of the generalized
Silin modes in a multivalley system, monolayer graphene.
A gauge- and Berry-gauge- invariant kinetic equation for a multivalley Fermi liquid is developed and applied to the case of graphene with extrinsic spin-orbit coupling (SOC).
The interplay of SOC and Berry curvature allows for the excitation of generalized Silin modes in the spin and valley-staggered-spin channels via an AC electric field.
The resonant contributions from these modes to the optical conductivity are calculated.
\end{abstract}

\maketitle

It has long been known that while
spin waves in a Fermi liquid are normally
overdamped~\cite{Lifsic2006}, in the presence of a finite Zeeman field
(a magnetic field acting only on particle spins)
there exist well-defined, gapped spin collective modes of the Fermi-liquid state, the Silin
modes~\cite{Silin1958,Platzman1967,Schultz1967,Leggett1970,Candela1986,Baboux2013,Baboux2015}.
In multi-valley materials there may be additional collective excitations of the Fermi liquid state, beyond the charge and spin modes.
These additional modes
are in general also diffusive in the time reversal symmetric case~\cite{Raines2021}.
However, in a finite Zeeman
field these modes may become oscillatory, generalizing the notion of the Silin mode.
The aim of this work is to identify these combined spin-valley modes in graphene and explore ways to excite them.

 We
 predict a set of generalized Silin modes in graphene, comprised of the spin density Silin mode as well as a new valley-staggered Silin mode, which become oscillatory in a finite in-plane magnetic field.
 The mode frequencies differ from each other due to the difference between the corresponding Landau Fermi-liquid constants.
 An effective coupling of an electromagnetic wave to the new valley-staggered modes can be achieved by engaging the spin-orbit coupling (SOC) induced by a substrate, see, e.g.,~\cite{Wang2016c}.
 Additionally, as is the case for the usual Silin mode, SOC allows the generalized Silin modes to be excited by the electric field of the wave; this results in electric dipole spin resonance (EDSR) being the dominant~\cite{Rashba1965,Rashba2003,Duckheim2006,Maiti2016} excitation mechanism.

 We investigate the structure and frequencies of generalized Silin modes
 and find the corresponding resonances in the conductivity tensor.
 The modes differ not only by their frequency, but also by sensitivity to the polarization of the EM wave exciting them, allowing modes of different character to be selectively excited.

We consider a graphene sheet with spin-orbit coupling induced by the substrate made, e.g., of a transition metal dichalcogenide (TMD)~\cite{Wang2016c}. In addition to SOC, the substrate in general also induces gaps at Dirac points in the graphene's electronic spectrum. We assume the electron density is tuned away from the charge-neutrality point, allowing us to apply the Fermi-liquid theory.
The system is subject to a static, in-plane magnetic field $\mathbf{H}_0$ needed for the generalized Silin modes, and an AC field $\mathbf{E}(t)$ probing them.

To describe the generalized Silin modes at long wavelength 
we augment the theory of a multi-valley Fermi liquid~\cite{Aleiner2007,Kharitonov2012,Raines2021} by including the effects of SOC and of the external fields.
To deduce the linear response to the probe fields, we must first obtain the energy functional with extrinsic spin-orbit coupling and applied Zeeman field.

\emph{Model.} In the presence of extrinsic spin-orbit coupling, the single-particle Dirac Hamiltonian written in the valley-sub-lattice basis $(KA, KB, K'B, -K'A)$ takes the form
\begin{equation}
\hat{H}_{\mathbf{p}} = v^{\phantom 2}_D \mathbf{p} \cdot \hat{\bm{\Sigma}} + \Delta \hat{\Sigma}_z\hat{\tau}_z + \lambda \hat{\tau}_z \hat{\sigma}_z + \lambda^{\phantom 2}_R  \mathbf{e}_z \cdot (\hat{\bm{\sigma}} \times \hat{\bm{\Sigma}})\,.
\label{eq:H0}
\end{equation}
Here $\hat{\bm{\Sigma}}$, $\hat{\bm{\tau}}$, and $\hat{\bm{\sigma}}$ are the vectors of Pauli matrices in the spaces of sub-lattices ($A,B$), points $K,K^\prime$ in the Brillouin zone, and electron spin, respectively, and 
$\mathbf{e}_z$ is the unit vector in the $z$ (out-of-plane) direction.
In the absence of SOC, the graphene spectrum is characterized by the Dirac velocity $v_D$ and gap $\Delta$; the valley-Zeeman ($\lambda$) and
Rashba ($\lambda_R$) spin-orbit couplings arise from the inversion symmetry breaking by and wave function hybridization with the TMD substrate~\cite{Wang2016c}.
For definiteness, we take the Fermi level to be in the upper band.
If the SOC couplings are small compared to the Fermi energy (as measured from charged neutrality), 
we may perform the projection onto the upper band perturbatively in $\lambda_R$ and $\lambda$, 
obtaining the effective single band Hamiltonian (see \cite{SM} for details)
\begin{equation}
\hat{H}^+_\mathbf{p} =
\epsilon_{\mathbf{p}}
- \frac{1}{2}\mu_s \mathbf{H}_0 \cdot \hat{\bm{\sigma}}
+ \alpha_R(p) (\mathbf{p} \times \mathbf{e}_z )\cdot \hat{\bm{\sigma}}+ \Lambda(p) \hat{\sigma}_z \hat{\tau}_z
\label{eq:HPlus}
\end{equation}
with $\epsilon_{\mathbf{p}} = \sqrt{v_D^2 p^2 + \Delta^2}$ being the massive Dirac dispersion, $\alpha_R(p) = v_D \lambda_R / \epsilon_\mathbf{p}$ the effective Rashba coupling, $\Lambda(p) = \lambda - \lambda_R^2 \Delta/\epsilon_{\mathbf{p}}^2$ the effecive valley-Zeeman coupling, and where we have included the Zeeman energy due to external magnetic field $\mathbf{H}_0$ for particles with
effective Bohr magneton
$\mu_s=g_s e/4m_e c$, where $g_s$ is the 
Land{\'e} factor and $m_e$ is the free electron mass.
A complete description of the dynamics of 
the projected upper 
band also requires the evaluation of the Berry connection, $\bm{\mathcal{\hat{A}}}$, which consists of Abelian and non-Abelian parts \cite{Culcer2005,Xiao2010}.
To leading order in the Rashba term, we find
$
\bm{\mathcal{\hat{A}}} = \bm{\mathcal{\hat{A}}}_0
+
\bm{\mathcal{\hat{A}}}_1 + \mathcal{O}\left(\alpha_R^3\right)$,
where
$\bm{\mathcal{\hat{A}}}_0$
  is the Abelian Berry connection of gapped graphene, while the non-Abelian part is given by
\begin{equation}
\bm{\mathcal{\hat{A}}}_1
= -\tilde{\alpha}_R(p)
2\pi \nu(\epsilon_{\mathbf{p}}) \abs{\Omega_0^z(p)} \hat{\bm{\sigma}}_\parallel\hat{\tau}_z
\label{eq:BerryConnection}.
\end{equation}

Here,
$\boldsymbol{\Omega}=\Omega^z_0 \mathbf{e}_z$ is the Berry curvature of gapped graphene, 
$
\Omega_0^z(p)
=\mp v_D^2\Delta/2\epsilon_{\mathbf{p}}^3$  
with $\mp$ corresponding to the $K$ ($K'$) point, 
$\nu(\epsilon) = \epsilon/2\pi v_D^2$ is the density of states of the graphene bands~\footnote{The Berry connection will only enter the collective mode equations of motion evaluated at the Fermi surface.
Thus in what follows, we set $p=p_F$ and suppress the momentum arguments.}, and $\hat{\bm{\sigma}}_\parallel=\hat \sigma_x \hat x+\hat\sigma_y\hat y$.
The tilde on $\tilde{\alpha}_R$ indicates renormalization of the effective Rashba strength which will be discussed below.
The valley-Zeeman term by itself does not give rise to a non-Abelian Berry connection because it commutes with the Dirac part of the Hamiltonian.

With \cref{eq:HPlus,eq:BerryConnection} we are able to write a kinetic equation for the projected upper band in the collisionless limit~\cite{Bettelheim2017}
\begin{equation}
  \partial_{t} \hat{\rho}+\frac{1}{2} \{\bm{\nabla}\hat{\rho}\overset{\cdot}{,} \hat{\mathbf{v}}
  \}
  + \frac{1}{2}\{\bm{\mathcal { D }}\hat{\rho}\overset{\cdot}{,} \hat{\bm{F}}\}
  +i\left[\hat{\epsilon},\hat{\rho}\right] =0,
\label{eq:bettelheim-eom}
\end{equation}
where $\hat \rho$ is the density matrix,  $\hat\epsilon$ is the (matrix) quasiparticle energy functional, $\hat{\bm{F}}$ is the total force (external plus self-consistent) acting on a quasiparticle,
and the Berry covariant derivative is defined as \cite{Culcer2005,Xiao2010}
\begin{equation}
\bm{\mathcal{D}} \hat{g}=\nabla^{(\mathbf{p})} \hat{g}-\imath[\bm{\mathcal{\hat{A}}}, \hat{g}]
\label{eq:long-deriv}
\end{equation}
with $\nabla^{(\mathbf{p})}$ 
denoting
the gradient in momentum space.
The velocity and force appearing in \cref{eq:bettelheim-eom} are governed by the quasiclassical equations of motion for the band~\cite{Bettelheim2017,Xiao2010}.
As we are working in the 2D limit, the non-Abelian Berry connection $\bm{\mathcal{A}}$ is completely in plane, while the non-abelian Berry curvature $\bm{\Omega}$ is entirely out of plane.
Here, we will take take $\mathbf{E},\mathbf{H}_0$ to be in plane, allowing the force and velocity terms to be simply written as
\begin{equation}
\hat{\mathbf{v}}\approx \bm{\mathcal{D}}\hat{\epsilon},\quad
\hat{\bm{F}}\approx e\bm{E} - \bm{\nabla}\hat{\epsilon}.\label{eq:v-and-F}
\end{equation}
Here we have noted that in \cref{eq:bettelheim-eom} $\mathbf{v}$ multiplies $\bm{\nabla}\hat{\rho}$. The latter appears in first order in $\mathbf{E}$, so within the linear response theory we can neglect terms of order $\mathbf{E}$ in $\mathbf{v}$.

The system of \cref{eq:bettelheim-eom,eq:v-and-F} provides a gauge- and Berry-gauge-invariant description of the dynamics of the system.
As such, it is a convenient launching point to incorporate the interplay between band topology, spin orbit coupling, and Fermi-liquid effects.


The quasiparticle energy functional is found by combining \cref{eq:HPlus} with the interactions allowed by the approximate $SU(2)$ spin and $U(1)$ valley symmetries of gapped graphene~\cite{Aleiner2007,Kharitonov2012,Raines2021}
\begin{multline}
\hat{\epsilon}_{\text{FL}}[\hat{\rho}_\mathbf{p}]=
\sum_{\mathbf{p}'} \left[
f^d_{\mathbf{p}\mathbf{p}'}  n_{\mathbf{p}'}
+ f^s_{\mathbf{p}\mathbf{p}'}  \mathbf{s}_{\mathbf{p}'}\cdot \hat{\bm{\sigma}}
+ f^{v\parallel}_{\mathbf{p}\mathbf{p}'}  \mathbf{Y}_{\mathbf{p}'} \cdot \hat{\bm{\tau}}_\parallel\right.\\
\left.+ f^{vz}_{\mathbf{p}\mathbf{p}'}  Y^z_{\mathbf{p}'}\hat{\tau}_z+ f^{m\parallel}_{\mathbf{p}\mathbf{p}'}  \mathbf{M}^i_{\mathbf{p}'} \cdot \hat{\bm{\sigma}} \hat{\tau}_{\parallel,i}
+ f^{mz}_{\mathbf{p}\mathbf{p}'}  \mathbf{M}^z_{\mathbf{p}'} \cdot \hat{\bm{\sigma}} \hat{\tau}_z
\right]\,,
   \label{eq:efl}
\end{multline}
where $\hat{{\tau}}_{\parallel,i}$ are the $\hat{\tau}_x$ and $\hat{\tau}_y$ components of $\hat{\bm{\tau}}$, and index $i$ is summed over $i=x,y$.
Here we have decomposed the density matrix in terms of symmetry distinguished channels,
\begin{equation}
\hat{\rho}_\mathbf{p} =
n_{\mathbf{p}} + \mathbf{s}_\mathbf{p} \cdot \hat{\bm{\sigma}}
+ \mathbf{Y}_\mathbf{p} \cdot \hat{\bm{\tau}} + \mathbf{M}_\mathbf{p}^i \cdot \hat{\bm{\sigma}} \hat{\tau}_i,
\label{eq:densitymatrix}
\end{equation}
and $f^\mu_{\mathbf{p}\mathbf{p'}}$ are the Landau-Fermi liquid interaction functions associated 
with the channel.
The collective variables in \cref{eq:densitymatrix} are the densities of: charge $n$, spin $\mathbf{s}$, valley pseudo-spin $\mathbf{Y}$, and spin-triplet valley pseudo-spin $\mathbf{M}^i$.

Before considering the collective modes, we must identify the equilibrium density matrix.
The equilibrium occupations in the spin and valley-spin channels are due to the presence of the external Zeeman field and Rashba coupling for the former, and valley-Zeeman coupling for the latter.
We consider the case where the thermal, spin-orbit, and magnetic energy scales are small compared to the Fermi energy with respect to the band edge, $T,\mu_s H_0, \lambda,\lambda_R\ll E_F - \Delta$.
%
In direct analogy with the standard computation of the spin magnetic moment of the Fermi liquid~\cite{Nozieres1999,Baym1991},
%
the equilibrium density matrix is given by \cref{eq:densitymatrix} with~\cite{SM}
\begin{equation}
\begin{gathered}
n_{\text{eq},\mathbf{p}} = n_F(\tilde{\epsilon}_\mathbf{p}),\quad
Y_{\text{eq},\mathbf{p}} = 0\\
   \mathbf{s}_{\text{eq},\mathbf{p}} = -\frac{\partial n_F}{\partial \epsilon} \left(\frac{1}{2}\tilde{\mu}_s\mathbf{H}_0- \tilde{\alpha}_R(p)\mathbf{p} \times \mathbf{e}_z\right),\\
\mathbf{M}^i_{\text{eq},\mathbf{p}} = \frac{\partial n_F}{\partial \epsilon} \tilde{\Lambda}(p)\mathbf{e}_z\delta_{iz}.
\end{gathered}
\label{eq:variables-eq}
\end{equation}
Here, $n_F$ is the Fermi function
of
the local excitation energy in the absence of SOC and magnetic field
given by
 $\tilde{\epsilon}_\mathbf{p} = \epsilon_\mathbf{p}
   + \sum_{\mathbf{p}'}
   f^{d}_{\mathbf{p}\mathbf{p}'} n_F(\tilde{\epsilon}_{\mathbf{p}'})$~\footnote{Only interactions in the density channel contribute to this term as in the absence of SOC and Zeeman fields $n_\mathbf{p}$ is the only non-zero collective coordinate in \cref{eq:efl}.},
and $\epsilon_{\mathbf{p}}$ is defined in \cref{eq:HPlus}.
$\tilde{\mu}_s$, $\tilde{\alpha}(p)$, and $\tilde{\Lambda}(p)$
are, respectively, the renormalized by interaction effective spin magnetic moment, 
Rashba SOC strength \cite{Shekhter2005}, and valley-Zeeman SOC strength:
\begin{equation}
\tilde{\mu}_s = \frac{\mu_s}{1 + F_0^s},\
\tilde{\alpha}_R(p) = \frac{\alpha_R(p)}{1 + F_1^s},\
\tilde\Lambda(p)= \frac{\Lambda(p)}{1 + F^{mz}_0}.
\label{eq:effectivemu}
\end{equation}
The Landau Fermi-liquid parameters in \cref{eq:effectivemu} are
\begin{equation}
F^\mu_l = G_s G_v \nu_F \int \frac{d\phi}{2\pi}\frac{d\phi'}{2\pi}
f^\mu(\mathbf{p}_F, \mathbf{p}'_F)e^{-il(\phi-\phi')},
\end{equation}
where $\phi,\phi'$ are the azimuths of the respective momenta on the Fermi circles in valleys $K$ and $K^\prime$ and $\mu=d,s,{v\!\!\parallel,} \, vz,m\!\!\parallel,mz$; also,
$\nu_F$ is the density of states at the Fermi surface, and $G_s$ and $G_v$ are the spin and valley degeneracy, respectively.

Following the above considerations, we
write the equilibrium energy
$\hat\epsilon_{\text{eq}}\equiv\hat{\epsilon}
[\hat{\rho}_\text{eq}]$
as
\begin{equation}
\hat{\epsilon}_\text{eq} =
\tilde{\epsilon}_\mathbf{p} - \left(\frac{1}{2}\tilde{\mu}_s \mathbf{H}_0 - \tilde{\alpha}_R(p)(\mathbf{p} \times \mathbf{e}_z)\right)\cdot \hat{\bm{\sigma}}
+ \tilde\Lambda\hat{\sigma}_z \hat{\tau}_z.
\label{eq:epsilon}
\end{equation}

\emph{Linear response.}
To find the EDSR response of the system, we need  to keep  only linear-order terms in the electric field.
We thus linearize the kinetic equation 
\eqref{eq:bettelheim-eom}
in the deviation from equilibrium
$\hat{\rho} \equiv \hat{\rho}_\text{eq} + \delta\hat{\rho}$:
\begin{multline}
\partial_t \delta\hat{\rho}
+
\frac{1}{2} \nabla \cdot \left[
\{\hat{\mathbf{v}}, \delta \hat{\rho} \}
- \{\delta\hat{\epsilon}
{,} \bm{\mathcal{D}}\hat{\rho}_\text{eq} \}\right]
+ i [\hat{\epsilon}_\text{eq}, \delta\hat{\bar\rho}]\\
=
- e\mathbf{E}\cdot \bm{\mathcal{D}} \hat{\rho}_\text{eq},
\label{eq:lineareom}
\end{multline}
where we have defined the first-order correction to the quasiparticle energy from fluctuations in terms of the the Fermi-liquid interactions
$\delta\hat{\epsilon} = \hat{\epsilon}_\text{FL}[\delta\hat{\rho}]$ (cf. Eq.~\eqref{eq:efl}),
and the local deviation from equilibrium
$
\delta\hat{\bar{\rho}} \equiv
\delta\hat{\rho} -(\partial n_F/\partial \epsilon) \delta \hat{\epsilon}$.

\begin{figure}
    \centering
    \includegraphics[width=\linewidth]{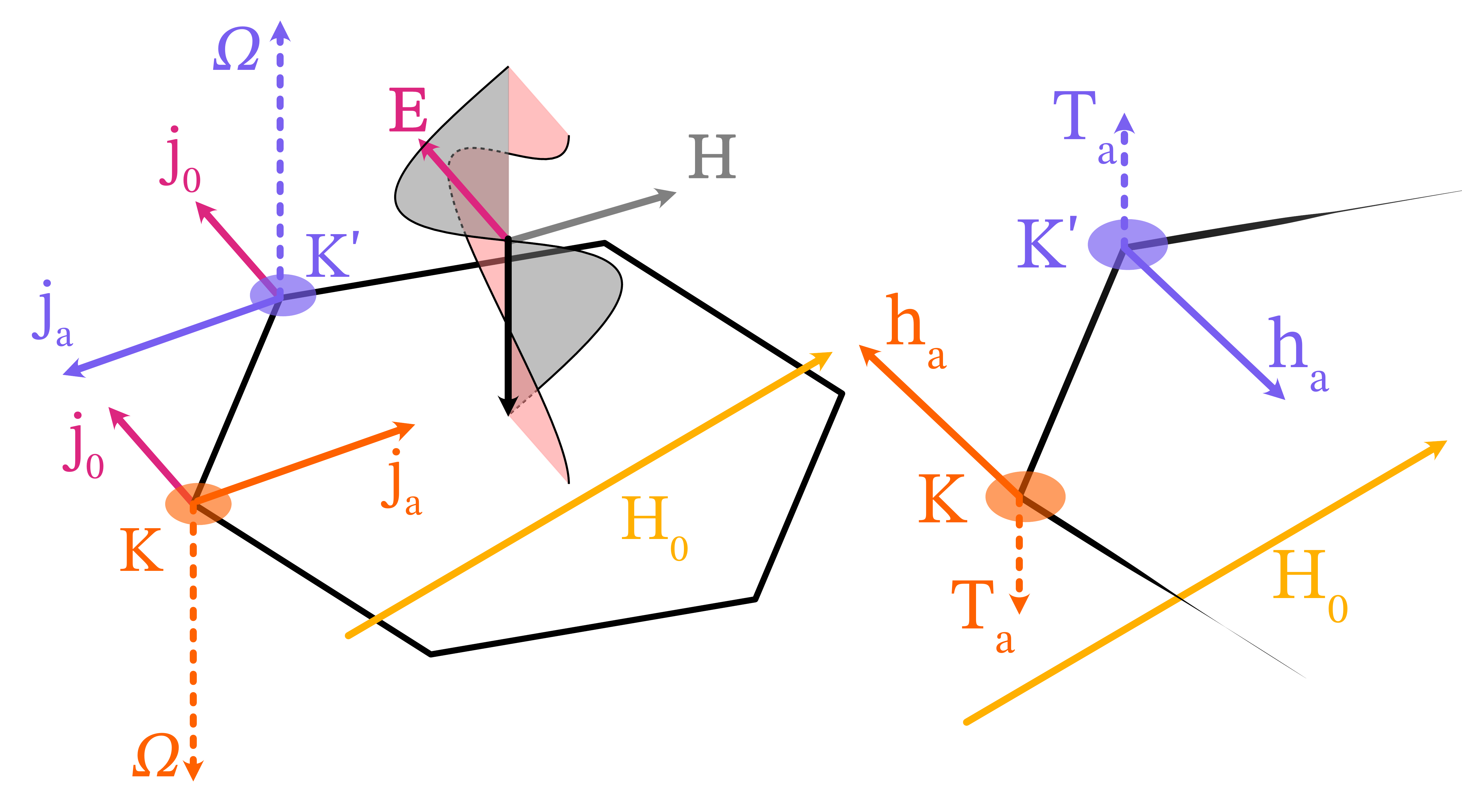}
    \caption{
    (Color online) \emph{Left}: Electron dipole spin resonance driven by an electromagnetic wave. The regular, $\mathbf{j}_0 \propto \mathbf{E}$, and anomalous, $\mathbf{j}_a \propto \bm{\Omega} \times \mathbf{E}$, currents at the $K$ (orange) and $K'$ (purple) points in the Brillouin zone, induced by
    the electric field $\mathbf{E}$ of the incident electromagnetic wave.
    Here, $\boldsymbol{\Omega}$ is the valley-staggered Berry curvature.
    \emph{Right}: Anomalous fields and torques.
    Spins are initially polarized along the Zeeman field $\mathbf{H}_0$.  The anomalous current-induced effective Rashba fields, $\mathbf{h}_a\propto \alpha_R\mathbf{e}_z \times \mathbf{j}_a$, 
    produce
    valley-specific 
    torques $\mathbf{T}_a \propto \mathbf{H}_0 \times \mathbf{h}_a$, 
    thus exciting the valley-staggered spin mode with an intensity proportional to $|\mathbf{E} \times \mathbf{H}|$.
    \label{fig:diagram}}
\end{figure}

From \cref{eq:lineareom} we obtain the conductivity as follows.
First, by taking the trace of \cref{eq:lineareom} and integrating it over momentum, we find the continuity equation,
\begin{equation}
 e \sum_\mathbf{p} \tr \partial_t\delta\hat{\rho} +
 \bm{\nabla} \cdot e\sum_\mathbf{p}\operatorname{tr}
\left(\hat{\mathbf{v}}_\mathbf{p} \delta\hat{\rho}
- \delta\hat{\epsilon}\bm{\mathcal{D}}\hat{\rho}_\text{eq}
\right)= 0\,,
\label{eq:continuity}
\end{equation}
which allows us to identify the longitudinal charge current as
\begin{equation}
\mathbf{j} =
e\sum_\mathbf{p}\operatorname{tr}
\left(\hat{\mathbf{v}}_\mathbf{p} \delta\hat{\rho}
- \delta\hat{\epsilon}\bm{\mathcal{D}}\hat{\rho}_\text{eq}
\right).
\label{eq:current1}
\end{equation}
Using the equilibrium energy c\ref{eq:epsilon} and density matrix of \cref{eq:variables-eq} allows us to write
the spin contribution to the longitudinal current (to order $\alpha_R^2$ and lowest order in $\lambda/\omega_s$).
%
Expanding the dynamic variables into
the angular harmonics on the Fermi surface,
\begin{equation}
\begin{gathered}
\delta\mathbf{s}_\mathbf{p} = -\frac{\partial n_F}{\partial \epsilon} \sum_l\delta\tilde{\mathbf{s}}_l e^{il\phi},\\
\delta\mathbf{M}^z_\mathbf{p} = -\frac{\partial n_F}{\partial \epsilon} \sum_l\delta\widetilde{\mathbf{M}}^z_l e^{il\phi},
\end{gathered}
\label{eq:angularmoments}
\end{equation}
we write the current as
\footnote{As with the Berry connection \cref{eq:BerryConnection} all functions of the magnitude of the momentum are here evaluated at $p=p_F$.
We have suppressed the momentum argument of such functions for compactness, e.g.~$\alpha_R = \alpha_R(p_F)$.}
\begin{multline}
  \mathbf{j}_\text{res}
  = -e G_s G_v \nu_F \tilde{\alpha}_R
  \left[\left(\frac{1}{2}\zeta +
2\pi\nu_F \abs{\Omega_0
^z}\frac{2\lambda}{\gamma_0}
  \right)\delta\tilde{\mathbf{s}}_{0} \times \mathbf{e}_z\right.
  \\
  \left.
  + \frac{1}{2}\left(1 + F_2^s - \frac{1}{2}\zeta \right)\sum_\pm \pm i \left(\mathbf{e}_{R/L} \cdot \delta\tilde{\mathbf{s}}_{\pm2}\right)\mathbf{e}_{R/L}\right]
  \\
  +2\pi e G_s G_v \nu_F^2\abs{\Omega_0^z} \tilde{\alpha}_R
  \left[\frac{1+F_0^{mz}}{1+F_0^s}\abs{\mu_s \mathbf{H}_0}
\delta\widetilde{M}^{
z
}_0
\mathbf{e}_y\right.
\\
+\left.\alpha_R p_F \left( \frac{1+F_1^s}{1+F_0^{mz}}
      +
      \frac{1}{2} \gamma_1
    \right)\sum_\pm \delta\widetilde{M}^{
    z
    }_{\pm1}{\mathbf{e}_{R/L}}\right],
  \label{eq:current}
\end{multline}
where $\mathbf{e}_{R/L} = \mathbf{e}_x \pm i \mathbf{e}_y$
and 
\begin{equation}
  \label{eq:zeta}
\zeta \equiv
 1 + \frac{E_F-\sqrt{E_F^2 - \Delta^2}}{E_F},\quad
 \gamma_l = \frac{1+F_l^{mz}}{1+F_l^s}.
\end{equation}
Note that the presence of $\Omega_0^z
$ in \cref{eq:current} indicates the important role of
the
Berry curvature in
driving
the
valley-staggered modes (see also \cref{fig:diagram}).

With an expression for the current in terms of $\delta\hat{\rho}$ we
can
solve the homogeneous $\mathbf{q}\to0$ limit of \cref{eq:lineareom} for $\delta\hat{\rho}$, obtaining a linear relationship between $\mathbf{j}$ and $\mathbf{E}$, 
and thus read off
the dissipative part of the optical conductivity tensor.

In the absence of
SOC
and
driving, the homogeneous limit of \cref{eq:lineareom} is simply
\begin{equation}
\partial_t \delta\hat{\rho}
+ i [\hat{\epsilon}_\text{eq}, \delta\hat{\bar\rho}]
= 0.
\label{eq:homogeneouseom}
\end{equation}
In the spin and valley-spin sectors, this gives the equations~\footnote{
Note we use the relations
$\delta{\tilde{\bar{\mathbf{s}}}}_l = (1+F_l^s)\delta{\tilde{\mathbf{s}}}_l $
,
$\delta{\widetilde{\overline{\mathbf{M}}}}^z_l = (1+F_l^{mz})\delta{\widetilde{\mathbf{M}}}^z_l$
to write \cref{eq:undriven-equations} entirely in terms of unbarred quantities.
The Fermi liquid parameters are absorbed into the definition of the mode frequencies $\omega_{sl},\omega_{ml}$.
}
\begin{equation}
\begin{gathered}
\partial_t \delta\tilde{\mathbf{s}}_l
- \omega_{sl}\mathbf{e}_x \times \delta\tilde{\mathbf{s}}_l
= 0, \quad \omega_{sl} = \frac{1 + F_l^s}{1+F_0^s} \abs{\mu_s H_0},\\
\partial_t \delta\widetilde{\mathbf{M}}^z_l
- \omega_{ml} \mathbf{e}_x \times \delta\widetilde{\mathbf{M}}^z_l
= 0, \quad \omega_{ml} = \gamma_l \omega_{sl}.
\end{gathered}
\label{eq:undriven-equations}
\end{equation}
Here $\delta\tilde{\mathbf{s}}$ corresponds to long wave-length spin-density modulations, while $\delta\widetilde{\mathbf{M}}^z$ describes spin-density modulations on length scale $|
\bm{K}-\bm{K}'|^{-1}$ corresponding to the separation between valleys in the Brillouin zone.
As in the case of the conventional Silin mode~\cite{Silin1958}, the mode frequencies are renormalized away from the Zeeman frequency by the Fermi-liquid
parameters,
with the exception of the $l=0$
mode frequency, which is protected by $SU(2)$ symmetry (see 
Ref.~\onlinecite{SM} for how this is modified by valley-Zeeman 
SOC).

\emph{Conductivity resonances.} Without 
SOC, the resonant finite frequency modes are not excited by an external electric field.
However, upon introduction of the Rashba coupling, three modes may be resonantly excited, namely, the $l=0$ spin
and valley-staggered spin modes,
and the $l=\pm2$ spin
mode.

The mechanism of driving can be understood as follows.
Initially, all particle spins are polarized along the external Zeeman field, which we take to be along the $\mathbf{e}_x$ axis.
Upon application of an external field $\mathbf{E}(t)$ the particle spins feel an effective magnetic field due to the Rashba term
$\mathbf{h} \propto \mathbf{e}_z \times \mathbf{j}$~\cite{Xiao2010,Pesin2012}
and therefore a spin torque~\cite{Manchon2009,Ado2017},
$
 \mathbf{T} \propto \mathbf{e}_x \times \mathbf{h} = \mathbf{e}_z j_x.
 $
The $x$ component of the current $j_x$ is composed of 
regular and anomalous pieces, shown in the left of \cref{fig:diagram},
\begin{equation}
j_x(\omega) = -
\frac{e^2 N E_x}{i\omega m^*} - e^2N\Omega_0^z
E_y,
\end{equation}
where $N$ is the number density, $m^*=p_F/v_F$ and $v_F=\epsilon'_{\mathbf{p}}\vert_{p=p_F}$.
The first of these terms creates identical torques
in both valleys, while the
second one, being proportional to the Berry curvature, yields valley-staggered torques depicted in the right of \cref{fig:diagram}.
Thus,
the component of $\mathbf{E}$ along $\mathbf{H}_0$
causes a valley-uniform torque on the spin, exciting the 
spin mode,
$\delta\tilde{\mathbf{s}}$,
 while
the component of $\mathbf{E}$ transverse to $\mathbf{H}_0$
causes a valley-staggered torque,
and thus excites the valley-staggered spin mode, $\delta\widetilde{\mathbf{M}}^z$.
Because the charge-to-spin conversion in both cases is proportional to the Rashba coupling, this leads to contributions to the conductivity proportional to $\alpha_R^2$.
Furthermore, the $\delta\tilde{\mathbf{s}}_0$ mode contributes to $\sigma^{xx}$ while the $\delta\widetilde{\mathbf{M}}^z_{0}$ mode contributes to $\sigma^{yy}$.

In the interacting case, the basic physical picture remains the same, but the quantities involved are renormalized.
The resonant contributions to the dissipative part of the conductivity can be written as (see Ref.~\onlinecite{SM})
\begin{equation}
\begin{aligned}
    \Re \sigma_1^{ii} =& \frac{1}{2}\pi e^2 G_s G_v \nu_F \tilde{\alpha}^2_R W_1^{ii}
    A_{m1}(\omega^2),\\
    \Re \sigma_0^{ii} =& \frac{1}{2}\pi e^2 G_s G_v \nu_F \tilde{\alpha}^2_R
    W_0^{ii}
    A_{s0}(\omega^2),\\
 \Re\sigma_2
  =& \frac{1}{2}\pi e^2 G_s G_v \nu_F \tilde{\alpha}^2_R
  W_2 A_{s2}(\omega^2)
\end{aligned}
\label{eq:conductivitieszero}
\end{equation}
where $A_{\mu l}(\omega^2) \equiv \omega_{\mu l} \delta(\omega^2 - \omega_{\mu l}^2)$ is the spectral function for the relevant mode and $W^{ii}_l$ is a dimensionless
peak weight ($i=x,y$).
Here the lack of indices on $\sigma_2$ indicates that the $l=2$ contribution is isotropic, $\sigma_2^{xx}=\sigma_2^{yy}=\sigma_2$,
and it is understood that the total conductivity  is the sum of the three lines in \cref{eq:conductivitieszero}.
Solving \cref{eq:lineareom} in the homogeneous limit and 
substituting the result into \cref{eq:current} we find, to order $\alpha_R^2$ and without the valley-Zeeman term,
\begin{equation}
\begin{gathered}
\lim_{\lambda\to0}    W_1^{ii}=0,\;\lim_{\lambda\to0}    W_0^{xx}=
    \zeta\frac{\omega_{s1}}{2\omega_{s0}},\\
\lim_{\lambda\to0}W^{yy}_{0}
=
\left(
  2\pi \nu_F \abs{\Omega_0^z}\right)^2\frac{\omega_{m0}^2}{1+F_0^{mz}}, \\
\lim_{\lambda\to0} W_2
  =
 2\left(1 + F_2^s - \frac{1}{2}\zeta \right)
 \left(1-\frac{\omega_{s1}}{2\omega_{s2}}\right).
\end{gathered}
\label{eq:weightszero}
\end{equation}
Here we explicitly see that the contribution from the $l=0$ modes has the strong anisotropy discussed above, while the $l=2$ contribution is isotropic.

Experiment shows that valley-Zeeman coupling is generally also present~\cite{Wakamura2018,Zihlmann2018,Wakamura2019}.
The inclusion of the valley-Zeeman coupling leads to a modification of the weights in \cref{eq:weightszero}.
Additionally, the valley-Zeeman term allows to excite the valley-staggered spin mode in the $l=1$ channel with frequency $\omega_{m1}$.
To lowest non-trivial order in the valley-Zeeman coupling $\lambda$, the weights in \cref{eq:conductivitieszero} are modified to
\begin{equation}
\begin{aligned}
W^{xx}_1
   =& 2\tilde{\lambda} (1+F_1^s)2\pi \nu_F \abs{\Omega_0^z}\\
   &\times
    \left(1+\frac{1}{2}\gamma_1^{-1}  \frac{1+F_1^s}{1+F_0^{mz}}
    \right)(1-\gamma_1),
\\
W^{yy}_{1}=&  2\tilde{\lambda}(1+F_1^s)2\pi\nu_F \abs{\Omega_0^z},  \\
&\times
\left[
\frac{\omega_{s1}-\omega_{m1}}{\omega_{m0}- \omega_{m1}}
+ (1-\gamma_1)\left(1 +\frac{1}{2}\gamma_1^{-1}  \frac{1+F_1^s}{1+F_0^{mz}}\right)\right]\\
W^{xx}_0 =&
\zeta\frac{\omega_{1s}}{2\omega_s}+
2\tilde{\lambda} (2\pi\nu_F \abs{\Omega_0^z})\left(1+F_1^s
-  \zeta\right)\\
W^{yy}_{0}
    = &
    2\pi \nu_F \abs{\Omega^z_0}
    \Bigg\{
      2\pi \nu_F \abs{\Omega_0^z}\frac{\omega^2_{m0}}{1+F_0^{mz}} \\
&+  2 \tilde{\lambda} \left[(1+F_1^{mz}) \frac{\omega_{s1} - \omega_{m0}}{\omega_{m0} - \omega_{m1}}
        -\frac{1}{1+F_0^{mz}} \zeta
\right]\Bigg\};
\end{aligned}
\label{eq:weightspert}
\end{equation}
while $W_2$ remains unchanged.
Note that there are two qualitatively different contributions to $W_1$.
One gives $W_1^{xx}$ and the second term in $W_2^{yy}$ (which are identical) and comes from coupling of the $l=1$ $\delta\widetilde{\mathbf{M}}^z$ mode to the $l=1$ spin
zero-mode $\delta\tilde{\mathbf{s}}_{l=\pm1}$, with magnetization parallel to
$\mathbf{H}_0$, via the valley-Zeeman SOC.
The other term, corresponding to the first term of $W_1^{yy}$ arises from conversion of the $l=0$ $\delta\widetilde{\mathbf{M}}^z$ mode into the $l=1$ mode via the Rashba SOC, which carries angular momentum $1$.
Both of these processes can only occur in the presence of interactions -- specifically when $F^{mz}_1 \neq F^s_1$ and $F^{mz}_1 \neq F^{mz}_0$ respectively~\footnote{The apparent divergence at $\omega_{m0}=\omega_{m1}$ is an artifact of the perturbation theory in $\lambda$ breaking down, as it is controlled by $\lambda/(\omega_{\mu l} - \omega_{\mu' l'})$} -- and arise due to the different effective magnetic moments for Zeeman vs valley-Zeeman fields.

\emph{Discussion.} It should be noted that these modes may be driven as well by an AC magnetic field.
Indeed, as discussed above, EDSR may be interpreted as being due to an effective Zeeman field created by the external electric field and Rashba coupling~\cite{Rashba1965,Maiti2016}.
The relative strength of EDSR driving compared to driving by an AC magnetic field 
is of the order of the ratio of atomic energy scale to driving frequency $E_\text{at}/\omega\gg 1$~\footnote{
The relative strength of the magnetic driving compared to the electric driving is determined by the ratio the magnetic and electric couplings~\cite{Shekhter2005}
$(e \alpha_R/\omega)/\mu_s
=([\{c/v_D\}\lambda_R]/\omega)(m_e v_D^2/g_sE_F)
.$
The SOC constant $\lambda_R$ is smaller than the characteristic atomic energy scale $E_\text{at}$ by the ratio of atomic velocity to speed of light, $\sim v_\text{at}/c$.
Therefore, the fraction $(c/v_D)\lambda_R/\omega$ is of the order $E_\text{at}/\omega\gg 1$
}, confirming the leading role of SOC in driving the modes~\cite{Rashba2003,Shekhter2005}.

The visibility of the generalized Silin modes in the optical conductivity will be determined by the broadening of the Silin mode peaks, as well as by the extent of the Drude peak tail.
In principle, this depends on two different relaxation times, the momentum relaxation time $\tau$ and spin relaxation time $\tau_s$.
We may approximate the effect of the spin relaxation on the Silin mode peaks by broadening the $\delta$-function peak to a Lorentizan of width $\tau_s^{-1}$.
Doing so, one may compute the ratio of the absorption peak height to the background Drude conductivity.
Writing the latter as
   $ \sigma_D(\omega) =
   e^2 G_s G_v \nu_F D/
   (1 - i\omega \tau)
   $
with diffusion constant $D = v_F^2\tau(1+F_0^d)/2,$ we can express the ratio of the resonant to Drude parts of the conductivity  as
\begin{equation}
\frac{\Re\sigma_\text{res}(\omega
=
\omega_i)}{\Re\sigma_D(\omega
=
\omega_i)}
\approx
\frac{1 + (\omega_i\tau)^2}{2}
\frac{\tilde{\alpha}^2_R}{v_F^2}\frac{\tau_s}{\tau} \frac{W_i}{1+ F_0^d}\,,
\end{equation}
where $W_i$ is the weight of the 
$\delta$-function for a resonant mode, cf. \cref{eq:weightspert,eq:conductivitieszero}.
The ratio $\tau_s/\tau_p$, extracted from weak anti-localization measurements
in graphene on TMD, varies between different studies~\cite{Wang2015a,Wang2016c,Wakamura2018,Zihlmann2018,Wakamura2019}. To be specific, we take
$\tau_s
\sim\tau
\sim\SI{1}{ps}$ \cite{Zihlmann2018}.
Then the resonant contribution is enhanced by applying a strong in-plane magnetic field ($H_0> \SI{10}{T}$) and also by choosing a material with larger $\tilde\alpha_R$.
From the beatings of Shubnikov-de Haas oscillations in bilayer graphene on WSe$_2$ one extracts $\lambda_R=10-15$\,meV  \cite{Wang2016c}; then $\tilde\alpha_R/v_F\sim \lambda_R/E_F\sim 0.1$.

To conclude, in this work we have shown that in the presence of an external magnetic field the normally diffusive spin-valley modes of graphene evolve into well-defined oscillatory modes with frequency set by the Larmor frequency, and Landau-Fermi liquid parameters, see Eq.~(\ref{eq:undriven-equations}).
The modes are a generalization of the Silin mode to multi-valley materials.
They can be probed via electric dipole spin resonance (EDSR) in the presence of extrinsic spin-orbit coupling.
Furthermore, certain modes may be selectively excited by changing the polarization of applied $E$ fields, leading to anisotropy of the optical conductivity, see Eqs.~(\ref{eq:conductivitieszero})-(\ref{eq:weightspert}).

\begin{acknowledgments}
Authors acknowledge discussions with H. Bouchiat, A. Kumar, S. Maiti, J. Meyer, O. Starykh, and T. Wakamura. This work was supported by NSF DMR-2002275 (LG), DMR-1720816 (DM), and the Yale Prize Postdoctoral Fellowship in Condensed Matter Theory (ZR). We 
acknowledge hospitality of KITP UCSB, supported by NSF PHY-1748958, (LG, DM) and LPS, University Paris-Sud, Orsay, France (DM).
\end{acknowledgments}

\bibliography{references.gen.bib}

\onecolumngrid
\renewcommand{\appendixpagename}{\centering Supplemental Material}
\setcounter{secnumdepth}{2}
\setcounter{section}{0}
\setcounter{equation}{0}
\setcounter{figure}{0}
\setcounter{table}{0}
\renewcommand{\thesection}{\Alph{section}}
\renewcommand{\thesubsection}{\thesection.\arabic{subsection}}
\appendixpage

\renewcommand{\theequation}{S\arabic{equation}}
\renewcommand{\thefigure}{S\arabic{figure}}
\renewcommand{\thetable}{S\arabic{table}}

\section{Evaluation of the equilibrium density matrix}
\label{sec:eqdensity}
We consider the case where the thermal, spin-orbit, and magnetic energy scales are small compared to the Fermi energy with respect to the band edge, i.e., $T,\mu_s H_0, \lambda,\lambda_R\ll E_F - \Delta$.
The equilibrium components of the quasiparticle energy functional in the spin and valley-spin channels are then obtained to lowest order in SOC and Zeeman field 
from the self-consistent equations
\begin{equation}
\begin{gathered}
\delta \varepsilon^s_\mathbf{p} = -\frac{1}{2} \mu_s \mathbf{H}_0 + \alpha_R(p) \mathbf{p} \times \mathbf{e}_z+ \sum_\mathbf{p'}  f^s_{\mathbf{p}\mathbf{p'}}
\frac{\partial n_{F}}{\partial \varepsilon}
\delta \varepsilon^s_\mathbf{p'}\,,\\
\delta \varepsilon^{mz}_\mathbf{p} = \Lambda(p) + \sum_\mathbf{p'}  f^{mz}_{\mathbf{p}\mathbf{p'}}
\frac{\partial n_{F}}{\partial \varepsilon}
\delta \varepsilon^{mz}_\mathbf{p'},
\end{gathered}
\label{eq:selfconsistent}
\end{equation}
in direct analogy with the standard computation of the spin magnetic moment of the Fermi liquid~\cite{Nozieres1999,Baym1991}.
To 
same order in SOC and the 
Zeeman field, the equilibrium density matrix can be obtained as 
linear response to new terms in the energy functional
\begin{equation}
\hat\rho_{\text{eq}}
\approx
n_F
(\tilde{\epsilon}) + \frac{\partial n_F}{\partial \epsilon} \left[
-\left(\frac{1}{2} \mu_s \mathbf{H}_0
- \alpha_R(p)\mathbf{p}\times \mathbf{e}_z
- \sum_{\mathbf{p}'} f^s_{\mathbf{p}\mathbf{p}'} \mathbf{s}_{\text{eq},\mathbf{p}'}
\right)\cdot \bm{\sigma}
+ \left(
\Lambda \mathbf{e}_z \delta_{i3} + \sum_{\mathbf{p}'} f^{mi}_{\mathbf{p}\mathbf{p}'} \mathbf{M}^i_{\text{eq},\mathbf{p}'}
\right)\cdot \bm{\sigma}\tau_i
\right]
.
\label{eq:eqrholinear}
\end{equation}
This motivates the parametrization
\begin{equation}
\mathbf{s}_{\text{eq}} = -\frac{\partial n}{\partial \epsilon} \left(\mathbf{m}- l \mathbf{p} \times \mathbf{e}_z\right),\qquad
\mathbf{M}^i_\text{eq} = \frac{\partial n}{\partial \epsilon} L \mathbf{e}_z\delta_{i3}.\label{eq:eqdefs}
\end{equation}
Plugging \cref{eq:eqdefs} 
into \cref{eq:eqrholinear} and using the fact that, at low $T$, the derivative of the Fermi function restricts momenta to the Fermi surface, we obtain the equations
\begin{equation}
\mathbf{m} - l \hat{p} \times \mathbf{e}_z
= \frac{1}{2}\mu_s \mathbf{H}_0 - F_0^s \mathbf{m} - \alpha_R(p) \mathbf{p}\times\mathbf{e}_z + l F_1^{s}\mathbf{p}\times\mathbf{e}_z,
\qquad
L = \Lambda - F_0^{mz}L.
\end{equation}
These may readily be solved for
\begin{equation}
\mathbf{m} = \frac{1}{2} \frac{\mu_s \mathbf{H}_0}{1 + F_0^s},\qquad l(p) = \frac{\alpha_R(p)}{1 + F_1^s},\qquad L = \frac{\Lambda}{1 + F^{mz}_0},\label{eq:eqvalues}
\end{equation}
Plugging these solutions back into $\hat{\epsilon}_0$ leads to \cref{eq:epsilon} with the renormalized constants
\begin{equation}
\tilde{\mu}_s = \mu_s \frac{1}{1 + F_0^s},\qquad \tilde{\alpha}_R(p) = \frac{\alpha_R(p)}{1 + F_1^s} = l(p),\qquad \tilde\lambda = \frac{\lambda}{1 + F^{mz}_0}= L,
\label{eq:renormcoupling}
\end{equation}
as they appear in the main text.

\section{Projection of the Hamiltonian onto the upper band}
\label{sec:secondorder}
In this section, we
derive the projected Hamiltonian
up to order $\lambda_R^2$,
employing the Schrieffer-Wolff transformation.
We start with
\begin{equation}
  \hat{H} = \hat{H}_D + \hat{H}_{VR},\quad
\hat{H}_D = \mathbf{d} \cdot \bm{\Sigma}, \quad \hat{H}_{VR} =  \lambda_R \mathbf{e}_z
\times \bm{\sigma}\cdot \bm{\Sigma}
\label{eq:Horig}
\end{equation}
with
\begin{equation}
\mathbf{d} = (v_D p_x, v_D p_y, \Delta\hat{\tau}_z)
\end{equation}
and $ \epsilon\equiv|\mathbf{d}|$, $\mathbf{n}\equiv \mathbf{d}/\epsilon$.
It is safe to neglect the valley-Zeeman term here as it
commutes
with $\hat{H}$.
We define the projectors on the upper/lower bands of $\hat{H}_D$ in \cref{eq:Horig} as
\begin{equation}
 \hat{P}_b = \frac{1}{2}\left(1 + b\mathbf{n} \cdot \bm{\Sigma}\right),
\end{equation}
and the Hamiltonians
\begin{equation}
 \hat{H}_R = \sum_b \hat{P}_b \hat{H}_{VR} \hat{P}_b, \quad \hat{H}_1 = \sum_b \hat{P}_b \hat{H}_{VR} \hat{P}_{-b}
 \label{eq:Hr}
\end{equation}
describing, respectively, the block-diagonal and block-off-diagonal components of the Rashba term.
The ``unperturbed'' block-diagonal Hamiltonian is thus given by
\begin{equation}
 \hat{H}_0 = \hat{H}_D + \hat{H}_R = \sum_b b\hat{P}_b \left(\epsilon + \alpha_R(p) \mathbf{e}_z \times \hat{\bm{\sigma}} \cdot \mathbf{p}\right), \quad \alpha_R = \frac{v_D \lambda_R}{\epsilon}.
 \label{eq:Hunperturbed}
\end{equation}
The total Hamiltonian is $\hat{H} = \hat{H}_0 + \hat{H}_1$, where $\hat{H}_1$ is the block-off-diagonal part.
Explicitly
\begin{equation}
  \hat{H}_1
  = \hat{H}_{VR} - \hat{H}_R
  = \lambda_R \mathbf{e}_z \times \hat{\bm{\sigma}} \cdot\left(\hat{\bm{\Sigma}} - \sum_b b\mathbf{n} \hat{P}_b\right).
  \label{eq:H1}
\end{equation}

\subsection{Canonical Transformation}
We now consider a transformation
\begin{equation}
  \label{eq:ut}
  \hat{U} = e^{\hat{T}},\quad \hat{T}^\dagger = - \hat{T},\quad \hat{P}_b \hat{T} \hat{P}_b = 0.
\end{equation}
Applying the transformation \cref{eq:ut} to \cref{eq:Horig} gives us a transformed Hamiltonian
\begin{equation}
  \label{eq:Hprime}
 \hat{H}' = \hat{U}\hat{H} \hat{U}^\dagger = \hat{U}\hat{H}_0\hat{U}^\dagger
+ \hat{U}\hat{H}_1\hat{U}^\dagger.
\end{equation}
Making use of the Baker-Campbell-Hausdorff relation, we expand $\hat{H}'$ to second order in $\hat T$
\begin{equation}
  \label{eq:HprimeT}
 \hat{H}' \approx \hat{H}_0 + \hat{H}_1 + [\hat{T}, \hat{H}_0] + [\hat{T}, \hat{H}_1] + \frac{1}{2} [\hat{T}, [\hat{T}, \hat{H}_0]] + \cdots
\end{equation}
The Schrieffer-Wolff transformation is effected by requiring that
\begin{equation}
  [\hat{T}, \hat{H}_0] = -\hat{H}_1 + \mathcal{O}(\lambda_R^3).
  \label{eq:Tcomm}
\end{equation}
Then to order $\lambda_R^2$, $\hat{H}'$ becomes block-diagonal:
\begin{equation}
  \label{eq:HprimeBD}
  \hat{H}' \approx \hat{H}_0
  + \frac{1}{2}[\hat{T}_1, \hat{H}_1] + \mathcal{O}(\lambda_R^3).
\end{equation}
Indeed, $\hat T$ can be always be chosen off-diagonal, while $\hat H_1$ is off-diagonal by construction, thus the product of $\hat T$ and $\hat H_1$ is diagonal.
Working at this order, it is straightforward to see that
if we choose
$\hat{T}$ to be
\begin{equation}
  \hat{T} = \sum_b b\hat{P}_b \frac{\hat{H}_1}{2\epsilon}\hat{P}_{-b} + \mathcal{O}(\lambda_R^3),
  \label{eq:Tdef}
\end{equation}
then \cref{eq:Tcomm} is indeed satisfied.
The second-order correction to the Hamiltonian  can be read off from \cref{eq:HprimeT,eq:Tcomm} as
\begin{equation}
 \hat{H}_2 = \frac{1}{2}[\hat{T},\hat{H}_1] .
 \label{eq:H2def}
\end{equation}
In particular, we 
are interested 
in the projection onto the upper band
\begin{equation}
 \hat{H}_2^+ \hat{P}_+ \equiv \hat{P}_+ \hat{H}_2 \hat{P}_+ .
\end{equation}
Noting that
\begin{equation}
  \hat{P}_+ \hat{T} = \frac{1}{2\epsilon} \hat{P}_+\hat{H}_1,
  \quad
  \hat{T}\hat{P}_+  = -\frac{1}{2\epsilon} \hat{H}_1\hat{P}_+
\end{equation}
we have
\begin{multline}
  \hat{H}_2^+\hat{P}_+ =
  \frac{1}{2\epsilon}\hat{P}_+ \hat{H}^2_2 \hat{P}_+
  =
  \frac{\lambda_R^2}{2\epsilon}\hat{P}_+
  {\left(
\mathbf{e}_z \times \hat{\bm{\sigma}}\cdot \left[\hat{\bm{\Sigma}} - \sum_b b \mathbf{n}\right]
    \right)}^2
  \hat{P}_+
  =
  \frac{\lambda_R^2}{2\epsilon}\hat{P}_+
  {\left(
\mathbf{e}_z \times \hat{\bm{\sigma}}\cdot \left[\hat{\bm{\Sigma}} - \mathbf{n}\right]
    \right)}^2
  \hat{P}_+ \\
  =
  \frac{\lambda_R^2}{2\epsilon}\hat{P}_+
\mathbf{e}_z \times \hat{\bm{\sigma}}\cdot \hat{\bm{\Sigma}}
  \left(
\mathbf{e}_z \times \hat{\bm{\sigma}}\cdot \left[\hat{\bm{\Sigma}} - \mathbf{n}\right]
    \right)
  \hat{P}_+
  = \frac{\lambda_R^2}{2\epsilon}\hat{P}_+
  \left[
  {\left(
\mathbf{e}_z \times \hat{\bm{\sigma}}\cdot \hat{\bm{\Sigma}}
\right)}^2
  - {\left(
\mathbf{e}_z \times \hat{\bm{\sigma}}\cdot \mathbf{n}
\right)}^2
\right]
  \hat{P}_+ \\
  = \frac{\lambda_R^2}{2\epsilon}
  \left[
\hat{P}_+
  {\left(
\mathbf{e}_z \times \hat{\bm{\sigma}}\cdot \hat{\bm{\Sigma}}
\right)}^2\hat{P}_+
  - \frac{v_D^2p^2}{\epsilon^2}
\hat{P}_+
\right].
\end{multline}
Evaluating
\begin{equation}
  {\left(
\mathbf{e}_z \times \hat{\bm{\sigma}}\cdot \hat{\bm{\Sigma}}
\right)}^2
=
  {\left(
      \hat{\sigma}_x \hat{\Sigma}_y
      - \hat{\sigma}_y \hat{\Sigma}_x
\right)}^2
= 1 + 1 - i\hat{\sigma}_z (-i\hat{\Sigma}_z) - (-i \hat{\sigma}_z) i \hat{\Sigma}_z
= 2 (1 - \hat{\sigma}_z \hat{\Sigma}_z),
\end{equation}
we then have
\begin{multline}
  \hat{H}_2^+
  = \frac{\lambda_R^2}{2\epsilon}
  \left[
    2(1 - \hat{\sigma}_z n_z)
  - \frac{v_D^2p^2}{\epsilon^2}
\right]
  = \frac{\lambda_R^2}{\epsilon}
  \left[
    1 - \frac{v_D^2p^2}{2\epsilon^2}
    - \frac{\Delta}{\epsilon}\hat{\sigma}_z \hat{\tau}_z
  \right]\\
  = \frac{\lambda_R^2}{\epsilon}
  \left[
    \frac{1}{2}\left(1 + \frac{\Delta^2}{\epsilon^2}\right)
    - \frac{\Delta}{\epsilon}\hat{\sigma}_z \hat{\tau}_z
  \right]
  = \frac{\lambda_R^2}{2\epsilon}
  {\left[
    1
    - \frac{\Delta}{\epsilon}\hat{\sigma}_z \hat{\tau}_z
  \right]}^2.
\end{multline}

\subsection{Berry connection}
Similarly, we may write the Berry connection as $\bm{\mathcal{\hat{A}}} = \bm{\mathcal{\hat{A}}}_0
+ \delta\bm{\mathcal{\hat{A}}}$,
where $\bm{\mathcal{\hat{A}}}_
0$ is the Berry connection associated with $\hat{H}_D$.
Again, We are interested in the upper band projection
\begin{equation}
\hat{P}_+\bm{\mathcal{\hat{A}}} \hat{P}_+ \equiv (\bm{\mathcal{\hat{A}}}_{
0+} + \delta\bm{\mathcal{\hat{A}}}_+)\hat{P}_+.
\end{equation}

Explicitly,
\begin{multline}
  \delta\bm{\mathcal{\hat{A}}}_+\hat{P}_+
  \equiv \hat{P}_+ \delta
  \bm{\mathcal{
  \hat{A}}}
  \hat{P}_+
  = i\hat{P}_+ \hat{U}(\bm{\nabla}^{(\mathbf{p})}
  \hat{U}^\dagger) \hat{P}_+
  \approx i\hat{P}_+
  \left(1 + \hat{T}\right)\bm{\nabla}^{(\mathbf{p})}\left(-\hat{T} + \frac{1}{2}\hat{T}^2\right)
  \hat{P}_+ \\
  = i\hat{P}_+
\left(-\bm{\nabla}^{(\mathbf{p})}\hat{T} +\frac{1}{2}\{\bm{\nabla}^{(\mathbf{p})} \hat{T}, \hat{T}\} -\hat{T}\bm{\nabla}^{(\mathbf{p})}\hat{T}\right)
  \hat{P}_+
  = -i\hat{P}_+
\left(\bm{\nabla}^{(\mathbf{p})}\hat{T} +\frac{1}{2}[\hat{T},\bm{\nabla}^{(\mathbf{p})}\hat{T}]\right)
  \hat{P}_+ .
\end{multline}
For the first term, we make use of \cref{eq:ut}
\begin{equation}
 \hat{P}_+(\bm{\nabla}^{(\mathbf{p})}\hat{T})\hat{P}_+
 =
 \cancelto{0}{\bm{\nabla}^{(\mathbf{p})}(\hat{P}_+\hat{T}\hat{P}_+)}
 - (\bm{\nabla}^{(\mathbf{p})}\hat{P}_+)\hat{T}\hat{P}_+
 - \hat{P}_+\hat{T}(\bm{\nabla}^{(\mathbf{p})}\hat{P}_+).
\end{equation}
So, at first order
\begin{multline}
  \delta\bm{\mathcal{\hat{A}}}_{1+}\hat{P}_+
  = \frac{i}{2}
(\bm{\nabla}^{(\mathbf{p})}\mathbf{n})\cdot
 \left[
 \hat{P}_+\hat{\bm{\Sigma}}\hat{P}_{-}\hat{T}\hat{P}_+
 + \hat{P}_+\hat{T}\hat{P}_{-}\hat{\bm{\Sigma}}\hat{P}_+
 \right]\\
  = \frac{i}{4\epsilon}
(\bm{\nabla}^{(\mathbf{p})}\mathbf{n})\cdot
 \left[
 \hat{P}_+\hat{\bm{\Sigma}}\hat{P}_{-}\hat{H}_1\hat{P}_+
 - \hat{P}_+\hat{H}_1\hat{P}_{-}\hat{\bm{\Sigma}}\hat{P}_+
 \right]\\
  = \frac{i}{4\epsilon}
(\bm{\nabla}^{(\mathbf{p})}\mathbf{n})\cdot
 \left[\hat{P}_+\hat{\bm{\Sigma}}\hat{H}_1\hat{P}_+
 - \hat{P}_+\hat{H}_1\hat{\bm{\Sigma}}\hat{P}_+
 \right]
  = \frac{i}{4\epsilon}
  \hat{P}_+ \left[
(\bm{\nabla}^{(\mathbf{p})}\mathbf{n})\cdot \hat{\bm{\Sigma}},\hat{H}_1
 \right]\hat{P}_+,
 \label{eq:A1pexpr}
\end{multline}
where
$[\bm{\nabla}^{(\mathbf{p})}\mathbf{n}]_{ij} 
\equiv\partial^p_i n_j$ 
and
we have used the idempotency of the upper band projection to sandwich \cref{eq:A1pexpr} with $\hat{P}_+$.
Using \cref{eq:H1},
\begin{multline}
  \delta\bm{\mathcal{\hat{A}}}_{1+}\hat{P}_+
  = \frac{i\lambda_R}{4\epsilon}
  \hat{P}_+ \left[
(\bm{\nabla}^{(\mathbf{p})}\mathbf{n})\cdot \hat{\bm{\Sigma}},
\mathbf{e}_z \times \hat{\bm{\sigma}} \cdot \left(\hat{\bm{\Sigma}} - \sum_b b \mathbf{n} \hat{P}_b\right)
 \right]\hat{P}_+
  = \frac{i\lambda_R}{4\epsilon}
  \hat{P}_+ \left[
(\bm{\nabla}^{(\mathbf{p})}\mathbf{n})\cdot \hat{\bm{\Sigma}},
\mathbf{e}_z \times \hat{\bm{\sigma}} \cdot \hat{\bm{\Sigma}}
 \right]\hat{P}_+\\
  = \frac{i\lambda_R}{4\epsilon}
  \hat{P}_+ 2i
(\bm{\nabla}^{(\mathbf{p})}\mathbf{n})\times
(\mathbf{e}_z \times \hat{\bm{\sigma}})
 \hat{P}_+
  = -\frac{\lambda_R}{2\epsilon}
(\bm{\nabla}^{(\mathbf{p})}\mathbf{n})\times
\left(\mathbf{e}_z \times \hat{\bm{\sigma}}\right) \cdot \mathbf{n} \hat{P}_+\\
  = -\frac{\lambda_R}{2\epsilon}
\left[(\bm{\nabla}^{(\mathbf{p})}\mathbf{n})\cdot \hat{\bm{\sigma}} \mathbf{e}_z
- (\bm{\nabla}^{(\mathbf{p})}\mathbf{n})\cdot
\mathbf{e}_z \hat{\bm{\sigma}}\right] \cdot \mathbf{n} \hat{P}_+.
\label{eq:A1p}
\end{multline}
Therefore,
\begin{equation}
  \delta\bm{\mathcal{\hat{A}}}_{1+}
  = \frac{\lambda_R}{2\epsilon}
\left[
(\bm{\nabla}^{(\mathbf{p})} n_z)
\mathbf{n} \cdot \hat{\bm{\sigma}}
-(\bm{\nabla}^{(\mathbf{p})}\mathbf{n})\cdot \hat{\bm{\sigma}} n_z
\right].
\label{eq:A1deriv}
\end{equation}
Now,
\begin{equation}
 \partial_i^p n_j = \frac{v_D\delta_{ij}}{\epsilon} - \frac{n_j v^2_D p_i}{\epsilon^3}
= \frac{v_D}{\epsilon}\left(\delta_{ij} - \frac{n_j v_D p_i}{\epsilon} \right)
= \frac{v_D}{\epsilon}\left(\delta_{ij} - n_i n_j \right).
\end{equation}
Plugging the last result back into \cref{eq:A1deriv}, we obtain 
\begin{equation}
  \delta\bm{\mathcal{\hat{A}}}_{1+}
  = \frac{\lambda_R}{2\epsilon}\frac{v_D}{\epsilon}
\left[
  -\mathbf{n} n_z
\mathbf{n} \cdot \hat{\bm{\sigma}}
-\hat{\bm{\sigma}} n_z
+\mathbf{n} \mathbf{n}\cdot \hat{\bm{\sigma}} n_z
\right]
  = -\frac{\lambda_R}{2\epsilon}\frac{v_D}{\epsilon}
\hat{\bm{\sigma}}_\parallel n_z
  = -\frac{\lambda_R}{2\epsilon}\frac{v_D}{\epsilon}
\hat{\bm{\sigma}}_\parallel n_z
  = -\frac{\lambda_R}{\epsilon}\frac{v_D\Delta}{2\epsilon^2}
\hat{\bm{\sigma}}_\parallel\hat{\tau}_z.
\label{eq:A1}
\end{equation}

Now for the second-order term we find
\begin{equation}
 \delta\bm{\mathcal{\hat{A}}}_2 \hat{P}_+
  = -\frac{i}{2}\hat{P}_+
[\hat{T},\bm{\nabla}^{(\mathbf{p})}\hat{T}]
  \hat{P}_+
  = -\frac{i}{4\epsilon}\hat{P}_+
\{\hat{H}_1,\bm{\nabla}^{(\mathbf{p})}\hat{T}\}
  \hat{P}_+
  = -\frac{i}{4\epsilon}\hat{P}_+
\bm{\nabla}^{(\mathbf{p})}\{\hat{H}_1,\hat{T}\}
  \hat{P}_+ .
\end{equation}
It follows from \cref{eq:Hr,eq:Tdef} that
\begin{equation}
  \{\hat{H}_1,\hat{T}\}
  = \sum_{bb'}
  \left[
    \hat{P}_b \hat{H}_{1} \hat{P}_{-b}
    \hat{P}_{b'} \frac{b'\hat{H}_{1}}{2\epsilon} \hat{P}_{-b'}
    +
    \hat{P}_{b} \frac{b\hat{H}_{1}}{2\epsilon} \hat{P}_{-b}
\hat{P}_{b'} \hat{H}_{1} \hat{P}_{-b'}
    \right]
  = \frac{1}{2\epsilon}\sum_{b}
\hat{P}_b
  \left[
    b-b
    \right]\hat{H}_1^2
\hat{P}_{b} = 0,
\end{equation}
and, therefore,  $\delta\bm{\mathcal{\hat{A}}}_+=\delta\bm{\mathcal{\hat{A}}}_{+1}$ to second order.
\subsection{Summary of the results}
To summarize, within second-order perturbation theory we have the effective upper band Hamiltonian
\begin{equation}
  \hat{H}^+ = \epsilon_\mathbf{p}
  + \alpha_R(p) \mathbf{p} \times \mathbf{e}_z \cdot \hat{\bm{\sigma}}
  + \frac{\lambda_R^2}{2\epsilon}
  {\left[
    1
    - \frac{\Delta}{\epsilon}\hat{\sigma}_z \hat{\tau}_z
  \right]}^2
 \label{eq:Hupper}
\end{equation}
with $\alpha_R = v_D \lambda_R /\epsilon$
and the effective Berry connection
\begin{equation}
  \bm{\mathcal{\hat{A}}} = \bm{\mathcal{\hat{A}}}_{
0+}+\delta\bm{\mathcal{\hat{A}}}_{
+}=\bm{\mathcal{\hat{A}}}_{
  0+} -\frac{\lambda_R}{\epsilon}\frac{v_D\Delta}{2\epsilon^2}
\hat{\bm{\sigma}}_\parallel\hat{\tau}_z
=
\bm{\mathcal{\hat{A}}}_{
0+} + \frac{\lambda_R}{v_D}\hat{\Omega}^z_0 \hat{\bm{\sigma}}_\parallel,
  \label{eq:Berry1}
\end{equation}
where $\Omega_0^z=-(v_D^2 \Delta/2\epsilon^3)\hat\tau_z$ is the Berry curvature near the gapped Dirac point.
\section{Conserved current}\label{sec:current-operator}

In 
the main text, we 
showed that the longitudinal current can be written as
\begin{equation}
\mathbf{j} =
e\sum_\mathbf{p}\operatorname{tr}
\left(\hat{\mathbf{v}}_\mathbf{p} \delta\hat{\rho}
- \delta\hat{\epsilon}\bm{\mathcal{D}}\hat{\rho}_\text{eq}
\right).
\label{eq:long-current}
\end{equation}
 It is 
 convenient
to break the current up into three pieces $\mathbf{j} = \mathbf{j}_\nabla +\mathbf{j}_{\mathcal{A}} + \mathbf{j}_\delta$, where
\begin{equation}
\mathbf{j}_\nabla =
e\sum_\mathbf{p}\operatorname{tr}
\bm{\nabla}^{(\mathbf{p})}\hat{\epsilon}_\text{eq} \delta\hat{\rho},\quad
\mathbf{j}_{\mathcal{A}} = -i e\sum_\mathbf{p}\operatorname{tr}
[\bm{\mathcal{\hat{A}}},
\hat{\rho}_\text{eq}]
\delta\hat{\bar \rho}
,\quad
\mathbf{j}_\delta
= -e\sum_\mathbf{p}\operatorname{tr} \delta\hat{\epsilon}\bm{\nabla}^{(\mathbf{p})}\hat{\rho}_\text{eq},
\label{eq:3j}
\end{equation}
where we have recalled that $\hat{\mathbf{v}}_\mathbf{p} = \bm{\nabla}^{(\mathbf{p})}\hat{\rho}_\text{eq} - i [\bm{\mathcal{\hat{A}}},
\hat{\rho}_\text{eq}]$
and have made use of the local deviation from equilibrium $\delta\hat{\bar{\rho}} \equiv
\delta\hat{\rho} -(\partial n_F/\partial \epsilon) \delta \hat{\epsilon}$.
In evaluating the current, we will use the parametrizations
\begin{equation}
\delta\hat{\rho} =
\sum_{\mathbf{p},l}
e^{i
l\phi}\left(-\frac{\partial n_F}{\partial \epsilon}\right)\left(
\delta n _l
+ \delta\tilde{\mathbf{s}}_l  \cdot \hat{\bm{\sigma}}
+ \delta\tilde{\mathbf{Y}}_l  \cdot \hat{\bm{\tau}}
+ \sum_{i=x,y,z}\delta\widetilde{\mathbf{M}}^i_l  \cdot \hat{\bm{\sigma}}\hat{\tau}_i
\right)
\end{equation}
and analogously for $\delta\hat{\bar{\rho}}$
\begin{equation}
\delta\hat{\bar\rho} =
\sum_{\mathbf{p},l}
e^{i
l\phi}\left(-\frac{\partial n_F}{\partial \epsilon}\right)\left(
\delta \bar{n} _l
+ \delta\tilde{\bar{\mathbf{s}}}_l  \cdot \hat{\bm{\sigma}}
+ \delta\tilde{\overline{\mathbf{Y}}}_l  \cdot \hat{\bm{\tau}}
+ \sum_{i=x,y,z}\delta\widetilde{\overline{\mathbf{M}}}^i_l  \cdot \hat{\bm{\sigma}}\hat{\tau}_i
\right),
\end{equation}
as well as the angular integrals
\begin{equation}
  \begin{gathered}
\Braket{
e^{i
l\phi}}_\text{FS}=\delta_{
l,0}\\
\Braket{
\mathbf{v}_F e^{
i
l\phi}}_\text{FS}
= \frac{1}{2}v_F  \mathbf{e}_{R/L} \delta_{
l,\pm1}
\\
\Braket{
v_{F,i}p_j e^{i
\phi}}_\text{FS}
= \frac{v_F p_F}{2}\delta_{
l,0}
+ \frac{v_F p_F}{4} e_{R/L, i} e_{R/L,j} \delta_{
l,\pm2}
\\
\Braket{
\bm{\nabla}^{
(
\mathbf{p}
)
}e^{i
l\phi}}_\text{FS}
=
\frac{1}{2p_F} \mathbf{e}_{R/L}\delta_{
l,\pm1}\\
\Braket{
(\nabla_i^{(\mathbf{p})}e^{i
l\phi})
 {\left[\mathbf{p} \times \mathbf{e}_z\right]}_j }_\text{FS}
= \pm i
\frac{1}{2}
e_{R/L,i}
e_{R/L,j}
\delta_{
l,\pm2}.
\end{gathered}
\label{eq:angularint}
\end{equation}
where $e_{R/L, i}$ is the $i$-th component of
\begin{equation}
  \label{eq:eRL}
\mathbf{e}_{R/L} = \mathbf{e}_x \pm i \mathbf{e}_y.
\end{equation}

\subsection{Gradient Term}
We begin with the gradient term
\begin{equation}
j^i_\nabla=
eG_s G_v \nu_F \sum_{\mathbf{p},l} \left\langle e^{i
l\phi}\left( v^i_F
\delta  n_l
+
\left.\frac{\partial \left(\tilde{\alpha}_R(p)\mathbf{p}\right)}{\partial p_i} \right|_{p_F}\times \mathbf{e}_z \cdot
\delta\tilde{{\mathbf{s}}}_l
+ \frac{\partial}{\partial p_i}\tilde{\Lambda}
\delta\widetilde{M}^{zz}_l
\right)
\right\rangle_{FS}
\end{equation}
Explicitly,
\begin{equation}
  \label{eq:alphaderiv}
\left.\frac{\partial \left(\tilde{\alpha}_R(p)\mathbf{p}\right)}{\partial p_i} \right|_{p_F}\times \mathbf{e}_z \cdot \delta\tilde{
\mathbf{s}
}_l
=
\left.\frac{\partial \left(\tilde{\alpha}_R(p)\mathbf{p}\right)}{\partial p_i} \right|_{p_F}\cdot \mathbf{e}_z \times \delta\tilde{
\mathbf{s}
}_l
=
\tilde{\alpha}_R(p_F)\left(\mathbf{e}_i
-\frac{v_F^i \mathbf{p}}{E_F}\right)  \cdot \mathbf{e}_z \times \delta\tilde{
\mathbf{s}
}_l.
\end{equation}
and
\begin{equation}
\bm{\nabla}^{(\mathbf{p})}\Lambda
= -\lambda_R^2\Delta \bm{\nabla}^{(\mathbf{p})}
\epsilon^{-2}
= 2 \lambda_R^2 \frac{\Delta}{2\epsilon^3} \mathbf{v}_F
= 2 \left(\frac{v_D \lambda_R}{\epsilon}\right)^2
\frac{v_D^2 \Delta}{2\epsilon^3} \frac{\epsilon^2}{v_D^4} \mathbf{v}_F
= 2 \alpha_R^2 (2\pi \nu_F) \abs{\Omega^z_0}\mathbf{p}_F,
\end{equation}
where we have used
\begin{equation}
    \mathbf{v}_F = \frac{v_D^2 \mathbf{p}_F}{E_F}.
\end{equation}
Writing
\begin{equation}
j^i_\nabla=
eG_s G_v \nu_F \sum_l \left( \braket{e^{il\phi}v^i_F}_{FS}
\delta  n_l
+
\Braket{e^{il\phi}\left.\frac{\partial \left(\tilde{\alpha}_R(p)\mathbf{p}\right)}{\partial p_i} \right|_{p_F}}\times \mathbf{e}_z \cdot
\delta {\mathbf{s}}_l
+ \Braket{e^{il\phi}\frac{\partial}{\partial p_i}\tilde{\Lambda}}
\delta\widetilde{M}^{zz}_l
\right)
\end{equation}
we may now use the relations \cref{eq:angularint}
\begin{multline}
  \mathbf{j}_\nabla
  = e G_s G_v \nu_F \left(
    \frac{v_F}{2} \sum_\pm \delta n_{\pm1}\mathbf{e}_{R/L}
  +  \frac{1}{1+F_0^{mz}}p_F \alpha_R^2 (2\pi \nu_F\abs{\Omega_0^z}) \sum_\pm \delta\widetilde{\mathbf{M}}^{zz}_{\pm1}\mathbf{e}_{R/L}\right.\\
\left.
+ \frac{1}{2}\zeta \tilde{\alpha}_R(p_F) \mathbf{e}_z \times \delta {\mathbf{s}}_{0}
- \tilde{\alpha}_R(p_F) \frac{v_F p_F}{4 E_F} \sum_\pm \pm i \mathbf{e}_{R/L} \left(\mathbf{e}_{R/L} \cdot \delta {\mathbf{s}}_{\pm2}\right)
\right)
  \label{eq:jgrad}
\end{multline}
where we used
$\mathbf{e}_z \times \mathbf{e}_{R/L} = \pm i \mathbf{e}_{R/L}$,
and defined
\begin{equation}
  \zeta \equiv 2\left(1 - \frac{v_F p_F}{2 E_F}\right) = 1 + \frac{E_F - \sqrt{E_F^2 - \Delta^2}}{E_F}.
\end{equation}

\subsection{Commutator term}
Let us first note 
that 
the zeroth-order term in \cref{{eq:Berry1}} for the Berry connection commutes with the energy functional
\begin{equation}
[\bm{\mathcal{\hat{A}}}_0, \hat{\epsilon}_\text{eq}] = 0,
\end{equation}
and thus we need only the commutator $[\delta\mathcal{\hat{A}}_+,\rho_{\text{eq}}]$.
Again note that the \emph{renormalized} quantities enter the Berry connection.
The commutator contribution to the current is then
\begin{equation}
  \mathbf{j}_{\mathcal{A}} = e G_s G_v \nu_F
\tilde{\alpha}_R
2\pi\nu_F\abs{\Omega_0^z}
  \left(
\omega_0\delta\widetilde{M}^{zz}_0
\mathbf{e}_y
+  2p_F \alpha_R \gamma_1 
\mathbf{e}_x
\delta\widetilde{M}^{zz}_0
+ 2\lambda \gamma_0^{-1} \mathbf{e}_z \times \mathbf{s}_0
\right),
\end{equation}
where we have defined
\begin{equation}
    \gamma_i = \frac{1+F_i^{mz}}{1+F_l^s}.
\end{equation}

\subsection{Interaction term}
For the last term in \cref{eq:3j} we have 
\begin{multline}
  \mathbf{j}_\delta =
  - \frac{1}{2} e G_s G_v \nu_F \sum_l
  \left(F_l^s \tilde{\delta\mathbf{s}}_l \cdot \left[-\frac{1}{2}\tilde{\mu}_s \mathbf{H}_0\right]\braket{\bm{\nabla}^{(\mathbf{p})}(e^{i
  l\phi})}_{FS}
    + F_l^s \tilde{\alpha}_R \braket{\mathbf{p}_F \times \mathbf{e}_z \cdot \tilde{\mathbf{s}}_l \bm{\nabla}^{(\mathbf{p})}(e^{i
    l\phi})}_{FS}\right.\\
 \left.
  + F_l^m \lambda\tilde{\alpha}_R 
  \braket{\bm{\nabla}^{(\mathbf{p})}(e^{i
  l\phi})}_{FS}
   \right)\\
   = - \frac{1}{2} e G_s G_v \nu_F \sum_l
  F_l^s \tilde{\alpha}_R \braket{\mathbf{p}_F \times 
  \mathbf{e}_z \cdot \tilde{\mathbf{s}}_l 
  \bm{\nabla}^{(\mathbf{p})}(e^{i
  l\phi})}_{FS}
  =- \frac{1}{2} e G_s G_v \nu_F F_2^s \tilde{\alpha}_R 
  \sum_\pm \pm i \mathbf{e}_{R/L} \mathbf{e}_{R/L} \cdot \delta\tilde{\mathbf{s}}_{\pm2}.
\end{multline}
For the resonant part of the current we thus have (to order $\lambda_R^2$)
\begin{multline}
  \mathbf{j}_\text{res}
  = -e G_s G_v \nu_F \tilde{\alpha}_R
  \left[\left(\frac{1}{2}\zeta +
2\pi\nu_F \abs{\Omega_p^z}\frac{2\lambda}{\gamma_0}
  \right)\delta\tilde{\mathbf{s}}_{0} \times \mathbf{e}_z
  + \frac{1}{2}\left(1 + F_2^s - \frac{1}{2}\zeta \right)\sum_\pm \pm i \left(\mathbf{e}_{R/L} \cdot \delta\tilde{\mathbf{s}}_{\pm2}\right)\mathbf{e}_{R/L}\right]
  \\
  +2\pi e G_s G_v \nu_F^2\abs{\Omega_0^z} \tilde{\alpha}_R
  \left[\frac{1+F_0^{mz}}{1+F_0^s}\abs{\mu_s \mathbf{H}_0}
\delta\widetilde{M}^{zz}_0
\mathbf{e}_y
+\alpha_R p_F \left( \frac{1+F_1^s}{1+F_0^{mz}}
      +
      \frac{1}{2} \gamma_1
    \right)\sum_\pm \delta\widetilde{M}^{zz}_{\pm1}{\mathbf{e}_{R/L}}\right].
    \label{eq:suprescurrent}
\end{multline}

\section{Eigenmodes in the presence of valley-Zeeman SOC}\label{sec:valley-zeeman}

In the presence of both valley-Zeeman 
and Rashba SOC,
the linearized equations of motion for the 
correction to the density matrix 
at zeroth order in $\alpha_R$ read
\begin{subequations}
\begin{gather}
\partial_t \delta\tilde{\mathbf{s}}^0_{\pm1}
+ \tilde{\mu}_s \mathbf{H}_0 \times \delta\tilde{
\bar{
\mathbf{s}}}
^0_{\pm1}
 -2\tilde{\lambda}\mathbf{e}_z \times \delta\widetilde{
 \overline{\mathbf{M}}}
 ^{z0}_{\pm1}
=
\frac{1}{2}\tilde{\mu}_s \mathbf{H}_0
\frac{1}{2p_F} e (E_x \mp i E_y)\label{eq:s0vz}\\
\partial_t \delta\widetilde{\mathbf{M}}^{z0}_{\pm1}
+ \tilde{\mu}_s \mathbf{H}_0 \times \delta\widetilde{\overline{\mathbf{M}}}^{z0}_{\pm1}
 -2\tilde{\lambda}\mathbf{e}_z \times \delta\tilde{\bar{\mathbf{s}}}^{0}_{\pm1}
=
-\tilde{\lambda} \mathbf{e}_z
\frac{1}{2p_F} e (E_x \mp i E_y).\label{eq:m0vz}
\end{gather}
\end{subequations}
To
first order in $\alpha_R$, the equations are
\begin{gather}
\partial_t \delta\tilde{\mathbf{s}}_0^1
+ \tilde{\mu}_s \mathbf{H}_0 \times \delta\tilde{
\bar{
\mathbf{s}}
}
_0^1
 -2\tilde{\lambda}\mathbf{e}_z \times \delta\widetilde{\overline{\mathbf{M}}}^{z1}_0
=
-\tilde{\alpha}_R 2\tilde{\lambda} R e\mathbf{E} \times \mathbf{e}_z
+ \tilde{\alpha}_R p_F
\sum_{\pm}
\pm i
\mathbf{e}_{R/L}
\times
\delta\tilde{\bar{\mathbf{s}}}^{0}_{\pm1}
\\
\partial_t \delta\widetilde{\mathbf{M}}_0^{z1}
+ \tilde{\mu}_s \mathbf{H}_0 \times \delta\widetilde{\overline{\mathbf{M}}}_0^{z1}
 -2\tilde{\lambda}\mathbf{e}_z \times \delta\tilde{\bar{\mathbf{s}}}^{1}_0
=
\tilde{\alpha}_R
R
(e \mathbf{E} \times
\tilde{\mu}_s \mathbf{H}_0)
+ \tilde{\alpha}_R p_F
\sum_{\pm}
\pm i
\mathbf{e}_{R/L}
\times
\delta\widetilde{\overline{\mathbf{M}}}^{z0}_{\pm1}\\
\partial_t \delta\tilde{\mathbf{s}}_{\pm2}^1
+ \tilde{\mu}_s \mathbf{H}_0 \times \delta\tilde{\bar{\mathbf{s}}}_{\pm2}^1
 -2\tilde{\lambda}\mathbf{e}_z \times \delta\widetilde{\overline{\mathbf{M}}}^{z1}_{\pm2}
=
\pm i\tilde{\alpha}_R
\left[
\frac{1}{2}
\left(e E_x \mp i e E_y\right)
\mathbf{e}_{L/R}
-
p_F
\mathbf{e}_{L/R}
\times
\delta\tilde{\bar{\mathbf{s}}}^0_{\pm1}
\right]\\
\partial_t \delta\widetilde{\mathbf{M}}_{\pm2}^{z1}
+ \tilde{\mu}_s \mathbf{H}_0 \times \delta\widetilde{\overline{\mathbf{M}}}^{z1}_{\pm2}
 -2\tilde{\lambda}\mathbf{e}_z \times \delta\tilde{\bar{\mathbf{s}}}^{1}_{\pm2}
=
i\tilde{\alpha}_R
p_F
\mathbf{e}_{L/R}
\times
\delta\widetilde{\overline{\mathbf{M}}}^{z0}_{\pm1}
\end{gather}
where we have defined
\begin{equation}
    R = 2\pi \nu_F \abs{\Omega_0^z}.
\end{equation}

\subsection{Without Rashba SOC and no driving}
In the limit $\alpha_R \to 0$ we can consider the undriven eigenmodes.
The equations of motion for the $l$-sector then become
\begin{gather}
\partial_t \delta\tilde{\mathbf{s}}_{l}
+ \tilde{\mu}_s \mathbf{H}_0 \times \delta\tilde{\bar{\mathbf{s}}}_{l}
 -2\tilde{\lambda}\mathbf{e}_z \times \delta\widetilde{\overline{\mathbf{M}}}^{z}_{l}
= 0\\
\partial_t \delta\widetilde{\mathbf{M}}^{z}_{l}
+ \tilde{\mu}_s \mathbf{H}_0 \times \delta\widetilde{\overline{\mathbf{M}}}^{z}_{l}
 -2\tilde{\lambda}\mathbf{e}_z \times \delta\tilde{\bar{\mathbf{s}}}_{l}
= 0.
\end{gather}
Defining
\begin{equation}
\gamma_l = \frac{1+F_l^{mz}}{1+F_l^s},\quad
\omega_{sl} = \frac{1+F_l^s}{1+F_0^s} \abs{\mu_s \mathbf{H}_0},\quad
\omega_{ml} = \gamma_l \omega_{sl}, \quad \lambda_l = \frac{\omega_{ml}}{\omega_{m0}}\lambda,
\label{eq:paramdefs}
\end{equation}
we write this as
\begin{gather}
\partial_t \delta\tilde{\mathbf{s}}_{l}
- \omega_{sl}\mathbf{e}_x\times \delta{\mathbf{s}}_{l}
 -2\lambda_l \mathbf{e}_z \times \delta{\mathbf{M}}^{z}_{l}
= 0\\
\partial_t \delta\widetilde{\mathbf{M}}^{z}_{l}
- \omega_{ml} \mathbf{e}_x\times \delta{\mathbf{M}}^{z}_{l}
 -2\lambda_l\gamma_l^{-1} \mathbf{e}_z \times \delta{\mathbf{s}}_{l}
= 0.
\end{gather}
This system of equations is simplified by defining
\begin{gather}
\mathbf{d}_{l,1} = (\sqrt{\gamma_l} \delta\widetilde{M}^{z}_x, \delta\tilde{s}_y, \delta\tilde{s}_z),\quad
\mathbf{d}_{l,2} = (\delta\tilde{s}_x, \sqrt{\gamma_l} \delta\widetilde{M}^{z}_y, \sqrt{\gamma_l} \delta\widetilde{M}^{z}_z),\label{eq:ddef}\\
\mathbf{h}_{l,1} = (\omega_{sl}, 0, 2\lambda_l \gamma_l^{-1/2}),\quad
\mathbf{h}_{l,2} = (\omega_{ml}, 0, 2\lambda_l \gamma_l^{-1/2})\label{eq:hdef}
\end{gather}
in which case the EOM is simply
\begin{equation}
    \partial_t \mathbf{d}_{l,i} - \mathbf{h}_{l,i} \times \mathbf{d}_{l,i} = 0.
\end{equation}
Thus in each $l$-sector we have two zero modes $\mathbf{d}_{l,i} \parallel \mathbf{h}_{l,i}$ and two finite frequency modes $\mathbf{d}_{l,i}\perp \mathbf{h}_{li,i}$ with
\begin{equation}
 \omega_{l,1} = \abs{h_{l,1}} = \sqrt{\omega_{sl}^2 + 4\lambda^2\gamma_l^{-1}},\quad
 \omega_{l,2} = \abs{h_{l,2}} = \sqrt{\omega_{ml}^2 + 4\lambda^2\gamma_l^{-1}}.
 \label{eq:omegalmu}
\end{equation}
These two modes can be seen to be adiabatically connected to the spin and valley-spin modes, respectively.

\section{Driving the modes}
In the presence of Rashba SOC there are two changes to the above: firstly, modes may now be driven by an external electric field, and secondly modes of different $l$ are coupled.

We treat this perturbatively in $\alpha_R$ as in the main text.

\subsection{Zeroth order in \texorpdfstring{$\alpha_R$}{aR}}
At zero-th order
\begin{equation}
    \partial_t \mathbf{d}_{\pm 1,i} - \mathbf{h}_{1,i} \times \mathbf{d}_{\pm 1,i} = \mathbf{f}_{\pm1, i}
\end{equation}
where
\begin{equation}
\mathbf{f}_{\pm1, 1} = 0,\quad
\mathbf{f}_{\pm1, 2} =
-\tilde{\omega}_s\mathbf{e}_x\mathcal{E}_\pm
-2\tilde{\lambda}\sqrt{\gamma_1}\mathbf{e}_z\mathcal{E}_\pm
= - \frac{1}{1+F_1^{s}}\mathbf{h}_{1,1}\mathcal{E}_\pm
 ,\quad
\mathcal{E}_\pm = \frac{1}{4 p_F} (e E_x \mp i e E_y).
\end{equation}
Thus we can see that the electric field drives
\begin{equation}
\mathbf{d}_{l=\pm 1, 2} = \left(
\frac{1}{i\omega} \mathbf{h}_{1, 1} \cdot \mathbf{h}_{1, 2}
+ \frac{1}{2} \sum_{\psi\in\{1,-1\}} \frac{i\mathbf{e}_{1,2,-\psi} \cdot \mathbf{h}_1}{\omega + i0 - \psi \omega_{1, 2}}
\right)\frac{\mathcal{E}_\pm}{1+F_1^s}
\end{equation}
where
\begin{equation}
\mathbf{e}_{l, i, \pm} = \mathbf{e}_y \pm i \frac{\mathbf{h}_{l,i} \times \mathbf{e}_y}{\omega_{l,i}}.
\label{eq:elpm}
\end{equation}
In this case
\begin{gather}
\mathbf{h}_{1,1} \cdot \mathbf{h}_{1,2} = \omega_{m1}^2 \gamma^{-1}_1 + 4\lambda_1^2\gamma_1^{-1} = \gamma_1^{-1} \left(\omega_{m1}^2 + 4\lambda_1^2\right)\\
\mathbf{h}_{1,1} \cdot \mathbf{h}_{1, 2} \times \mathbf{e}_y
= \mathbf{h}_{1,1} \times \mathbf{h}_{1, 2} \cdot \mathbf{e}_y
= 2\lambda_1\gamma_1^{-1/2}\left(\omega_{m1} - \omega_{s1}\right)
\end{gather}
giving
\begin{equation}
\mathbf{d}_{l=\pm1,1} = 0,\quad
\mathbf{d}_{l=\pm 1, 2} = \left(
\frac{\omega_{m1}^2 + 4\lambda_1^2}{i\gamma_1\omega}
- \frac{2\lambda_1 \gamma_1^{-1/2}(\omega_{m1} - \omega_{s1})}{\omega_{1,2}}\frac{1}{2} \sum_{\psi\in\{1,-1\}} \frac{i\psi}{\omega + i0 - \psi \omega_{1,2}}
\right)\frac{\mathcal{E}_\pm}{1+F_1^s}.
\label{eq:d1}
\end{equation}
It is convenient to write this as $\mathbf{d}_{l=\pm, 1(2)} = \bm{\chi}_{1(2)} \frac{\mathcal{E}_\pm}{1+F_1^s}$
with
\begin{equation}
\bm{\chi}_1=0,\quad
\bm{\chi}_{2} =
\frac{\omega_{m1}^2 + 4\lambda_1^2}{i\gamma_1\omega}
- \frac{2\lambda_1 \gamma_1^{-1/2}(\omega_{m1} - \omega_{s1})}{\omega_{1,2}}\frac{1}{2} \sum_{\pm} \frac{i\psi}{\omega + i0 \mp\omega_{1,2}}.
\label{eq:chi12}
\end{equation}

\subsection{First order in \texorpdfstring{$\alpha_R$}{aR}, l=0}

At first order, we have for the $l=0$ equations
\begin{equation}
    \partial_t \mathbf{d}_{0,i} - \mathbf{h}_{0,i} \times \mathbf{d}_{0,i} = \mathbf{f}_{0,i}.
\end{equation}
To obtain the RHS we start by rewriting
\begin{equation}
\end{equation}
\begin{gather}
\partial_t \delta\tilde{\mathbf{s}}_0^1
+ \omega_s \mathbf{e}_x \times \delta\tilde{\bar{\mathbf{s}}}_0^1
 -2\lambda\mathbf{e}_z \times \delta\widetilde{\mathbf{M}}^{z1}_0
=
-\tilde{\alpha}_R 2\tilde{\lambda} R e\mathbf{E} \times \mathbf{e}_z
+ \tilde{\alpha}_R p_F
\sum_{\pm}
\pm i
\mathbf{e}_{R/L}
\times
\mathbf{e}_x \chi_{2x} \mathcal{E}_\pm
\\
\partial_t \delta\widetilde{\mathbf{M}}_0^{z1}
+ \omega_{m0} \mathbf{e}_x \times \delta\widetilde{\mathbf{M}}_0^{z1}
 -\frac{2\lambda}{\gamma_0} \mathbf{e}_z \times \delta\tilde{\mathbf{s}}^{1}_0
=
\tilde{\alpha}_R
R
(e \mathbf{E} \times
\tilde{\mu}_s \mathbf{H}_0)
+ \tilde{\alpha}_R p_F
\sum_{\pm}
\pm i
\mathbf{e}_{R/L}
\times
\left(\chi_{2y} \mathbf{e}_y + \chi_{2z} \mathbf{e}_z\right)\sqrt{\gamma_1}\mathcal{E}_\pm.
\end{gather}
Performing the sums on the RHS gives
\begin{gather}
\partial_t \delta\tilde{\mathbf{s}}_0^1
+ \omega_s \mathbf{e}_x \times \delta\tilde{\bar{\mathbf{s}}}_0^1
 -2\lambda\mathbf{e}_z \times \delta\widetilde{\mathbf{M}}^{z1}_0
=
-\tilde{\alpha}_R 2\tilde{\lambda} R e\mathbf{E} \times \mathbf{e}_z
+ \frac{1}{2}\tilde{\alpha}_R \mathbf{e}_z
\chi_{2x} e E_x
\\
\partial_t \delta\widetilde{\mathbf{M}}_0^{z1}
+ \omega_{m0} \mathbf{e}_x \times \delta\widetilde{\mathbf{M}}_0^{z1}
 -\frac{2\lambda}{\gamma_0} \mathbf{e}_z \times \delta\tilde{\mathbf{s}}^{1}_0
=
\tilde{\alpha}_R
R
e E_y \tilde{\omega}_s \mathbf{e}_z
+ \frac{1}{2}\tilde{\alpha}_R
\left(\chi_{2y} \mathbf{e}_z  eE_y
- \chi_{2z} \mathbf{e}_y e E_y
- \chi_{2z} \mathbf{e}_x e E_x
\right)\sqrt{\gamma_1}
\end{gather}
Performing the change of basis to the $d$ modes we thus find
\begin{equation}
\begin{gathered}
\mathbf{f}_{0,1}
= e E_x\tilde{\alpha}_R\left(-\frac{1}{2}\sqrt{\gamma_0\gamma_1}\mathbf{e}_x
+ 2\tilde{\lambda} R \mathbf{e}_y
+ \frac{1}{2} \chi_{2x}\mathbf{e}_z\right),\\
\mathbf{f}_{0, 2} =
e E_y\tilde{\alpha}_R  \left(
-2\tilde{\lambda}  R \mathbf{e}_x
+
\frac{1}{2}\sqrt{\gamma_0\gamma_1}\chi_{2y} \mathbf{e}_z
- \frac{1}{2}\sqrt{\gamma_0\gamma_1}\chi_{2z} \mathbf{e}_y
\right).
\end{gathered}
\label{eq:f0}
\end{equation}
Focusing on the finite frequency modes we may thus solve, in the same way as for $l=0$,
\begin{equation}
\mathbf{d}_{0,i,\text{res}} = \frac{1}{2}\sum_\pm \mathbf{e}_{0,i,\pm} \frac{i \mathbf{e}_{0, i,\mp} \cdot \mathbf{f}_{0,i}}{\omega + i0 \mp \omega_{0,i}}
\label{eq:d0}
\end{equation}
where $h$ is defined as in \cref{eq:hdef} and $\mathbf{e}_{l,i,\pm}$ as in \cref{eq:elpm}.

\subsection{First order in \texorpdfstring{$\alpha_R$}{aR}, l=2}

Similarly for the $l=2$ mode the equation of motion we must find $\mathbf{f}_{\pm2,i}$.
We start again with the equations of motion
\begin{gather}
\partial_t \delta\tilde{\mathbf{s}}_{\pm2}^1
- \omega_{s2}\mathbf{e}_x \times \delta{\mathbf{s}}_{\pm2}^1
 -2\lambda_2\mathbf{e}_z \times \delta{\mathbf{M}}^{z1}_{\pm2}
=
\pm i\frac{1}{2}\tilde{\alpha}_R
\left[
\mathbf{e}_{L/R}
\mp \frac{1}{2}
i \mathbf{e}_{z}
\chi_{2x}
\right]
\left(e E_x \mp i e E_y\right)
\\
\partial_t \delta\mathbf{M}_{\pm2}^{z1}
- \omega_{m2} \mathbf{e}_x \times \delta{\mathbf{M}}^{z1}_{\pm2}
 -\frac{2\lambda_2}{\gamma_2}\mathbf{e}_z \times \delta{\mathbf{s}}^{1}_{\pm2}
=
i\tilde{\alpha}_R
\frac{1}{4}\sqrt{\gamma_1}
\left(
\chi_{2z}\mathbf{e}_z
- \chi_{2z} \mathbf{e}_y
\mp i\chi_{2z} \mathbf{e}_x
\right)
\left(e E_x \mp i e E_y\right).
\end{gather}
Thus
\begin{equation}
\begin{gathered}
 \mathbf{f}_{\pm2,1} =
 \frac{1}{2}\tilde{\alpha}_R
\left(
\pm\frac{1}{2}\sqrt{\gamma_2\gamma_1}
\chi_{2z} \mathbf{e}_x
-
\mathbf{e}_{y}
+\frac{1}{2}
\mathbf{e}_{z}
\chi_{2x}
\right)
\left(e E_x \mp i e E_y\right)\\
 \mathbf{f}_{\pm2,2} =
\pm i\frac{1}{2}\tilde{\alpha}_R \left(
\mathbf{e}_x
+
\frac{1}{2}\sqrt{\gamma_0\gamma_1}
\left[
\chi_{2z}\mathbf{e}_z
- \chi_{2z} \mathbf{e}_y
\right]
\right)
\left(e E_x \mp i e E_y\right).
\end{gathered}
\label{eq:f2}
\end{equation}
Focusing on the finite frequency modes we may thus solve, in the same way as for $l=\pm2$,
\begin{equation}
\mathbf{d}_{\pm2,i,\text{res}} = \frac{1}{2}\sum_{\psi \in \{1,-1\}} \mathbf{e}_{\pm2,i,\psi} \frac{i \mathbf{e}_{\pm2, i,-\psi} \cdot \mathbf{f}_{\pm2,i}}{\omega + i0 -\psi \omega_{\pm2,i}}
\label{eq:d2}
\end{equation}
where $h$ is defined as in \cref{eq:hdef} and $\mathbf{e}_{l,i,\pm}$ as in \cref{eq:elpm}.

\subsection{Summary}
The solutions \cref{eq:d1,eq:chi12,eq:d0,eq:f0,eq:d2,eq:f2}, along with the definitions \cref{eq:paramdefs,eq:ddef,eq:hdef,eq:omegalmu}, may be plugged into \cref{eq:suprescurrent} to obtain an expression for the optical conductivity for an arbitrary ratio of $\lambda_l/\omega_{sl}$.

\end{document}